\documentclass[aps,prl,twocolumn,superscriptaddress]{revtex4-1}

\usepackage{graphicx}
\usepackage{amssymb}
\usepackage{amsmath}
\usepackage{color}
\usepackage{physics}
\usepackage{bbm}
\usepackage{hyperref}
\usepackage{epstopdf}
\usepackage{pdfpages}

\renewcommand{\b}{\mathbf}

\makeatletter
\AtBeginDocument{\let\LS@rot\@undefined}
\makeatother

\begin{document}
\title{Floquet Hopf Insulators}

\author{Thomas Schuster}
\affiliation{Department of Physics, University of California, Berkeley, California 94720 USA}
\author{Snir Gazit}
\affiliation{Racah Institute of Physics and the Fritz Haber Center for Molecular Dynamics, The Hebrew University, Jerusalem 91904, Israel}
\affiliation{Department of Physics, University of California, Berkeley, California 94720 USA} 
\author{Joel E. Moore}
\affiliation{Department of Physics, University of California, Berkeley, California 94720 USA}
\affiliation{Materials Science Division, Lawrence Berkeley National Laboratory, Berkeley, California 94720, USA}
\author{Norman Y. Yao}
\affiliation{Department of Physics, University of California, Berkeley, California 94720 USA}
\affiliation{Materials Science Division, Lawrence Berkeley National Laboratory, Berkeley, California 94720, USA}
\date{\today}


\newcommand{\FigureDecomposition}{
\begin{figure}
\centering
\includegraphics[width=\columnwidth]{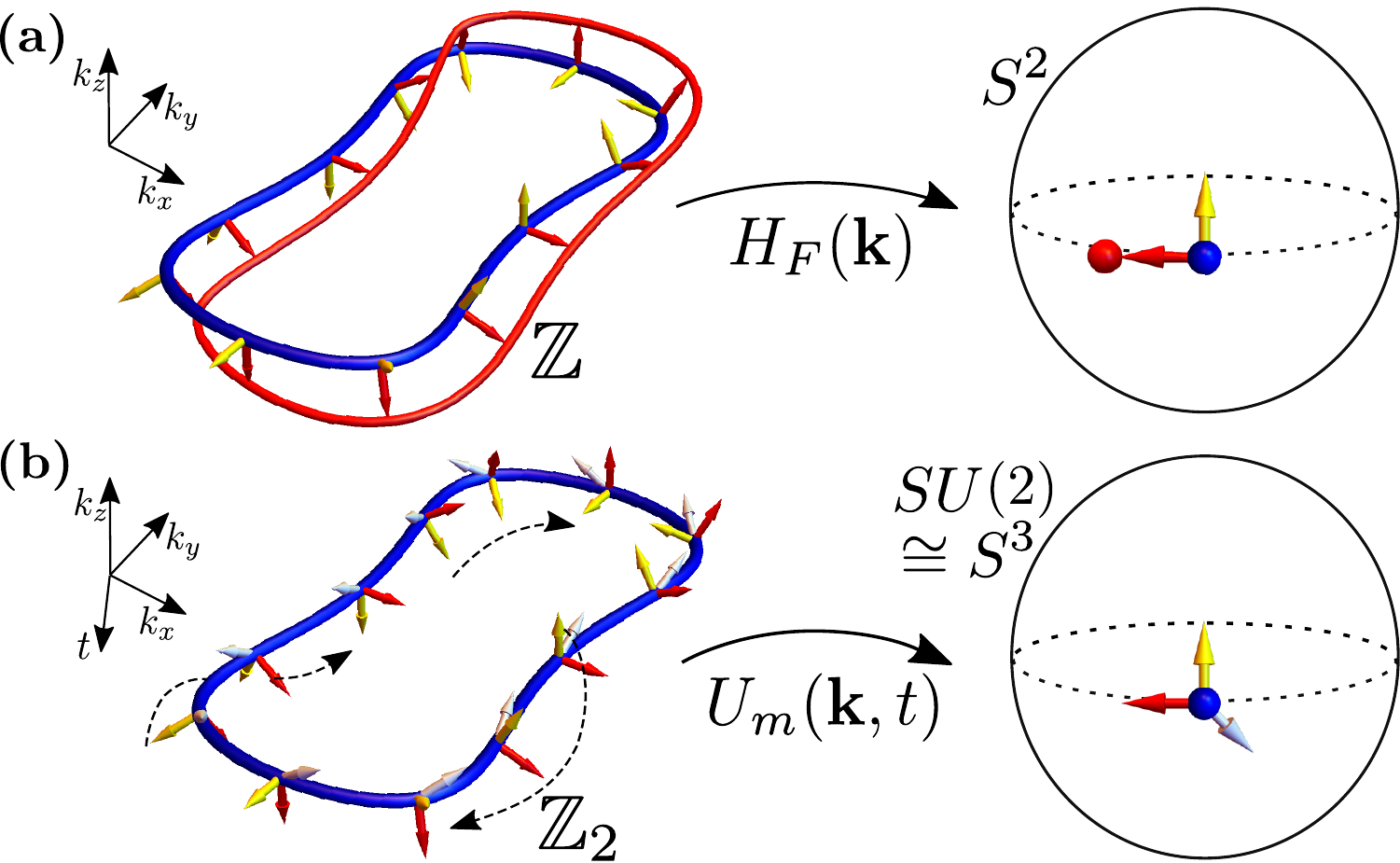}
\caption{
Depiction of the Floquet Hopf insulator's two topological invariants. 
\textbf{(a)} The `static' $\mathbb{Z}$ invariant is the Hopf invariant of the Floquet Hamiltonian $H_F(\b{k})$, corresponding to the linking of the pre-images (blue, red) of two points on the Bloch sphere. For nearby points, this equals the twisting of the Jacobian (colored arrows) along a single pre-image. 
\textbf{(b)}  The `Floquet' $\mathbb{Z}_2$ invariant classifies the micromotion operator $U_m(\b{k},t) \in SU(2)$, and is similarly interpreted as the Jacobian twisting (dashed black arrows) along a pre-image, with a reduced classification due to the larger dimensionality.} 
\label{fig: decomposition}
\end{figure}
}

\newcommand{\FigureHomotopy}{
\begin{figure*}
\centering
\includegraphics[width=\textwidth]{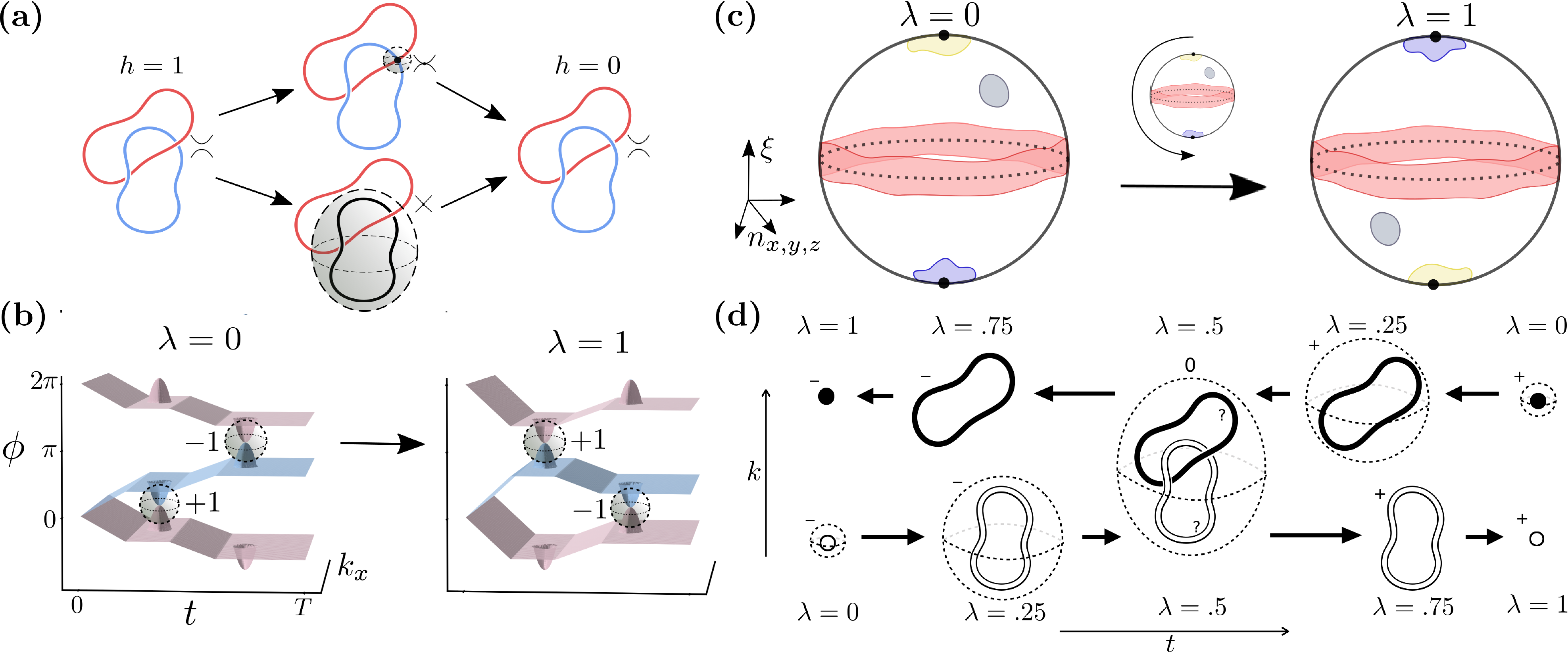}
\caption{
\textbf{(a)} Hopf defects are topologically protected gapless regions that the change the instantaneous Hopf invariant. A point Hopf defect (black point) has quadratic dispersion, and appears as a strand crossing that changes the linking number of any two eigenvectors' pre-images (red and blue). A loop Hopf defect (black loop) has linear dispersion, and can occur along a former pre-image. The defect charge is defined on a surface (gray, shaded) enclosing the defect.
\textbf{(b)} Schematic of two Floquet evolutions with different defect charges but the same topological invariants, which are connected by a smooth deformation $\lambda \in [0,1]$ that preserves the Floquet unitary's band gaps.
\textbf{(c)} The deformation can be viewed as a $\pi$ rotation of the 3-sphere parameterized by $(\b{n},\xi)$. Images of time-slices representing the initial $0$-defect (yellow), $\pi$-defect (blue), trivial Hopf invariant (gray), Hopf invariant $1$ (red), are displayed before and after the rotation.
\textbf{(d)} During the deformation, the $0$-defects (black outline) and $\pi$-defects (solid black) become loops that \emph{link} in the Brillouin zone, at which point their individual charges are undefined and may change. The total charge $h_0 + h_\pi$ is conserved and corresponds to the static $\mathbb{Z}$ invariant. Arrows indicate increasing $\lambda$.
} 
\label{fig: homotopy}
\end{figure*}
}

\newcommand{\FigureNumerics}{
\begin{figure}
\centering
\includegraphics[width=\columnwidth]{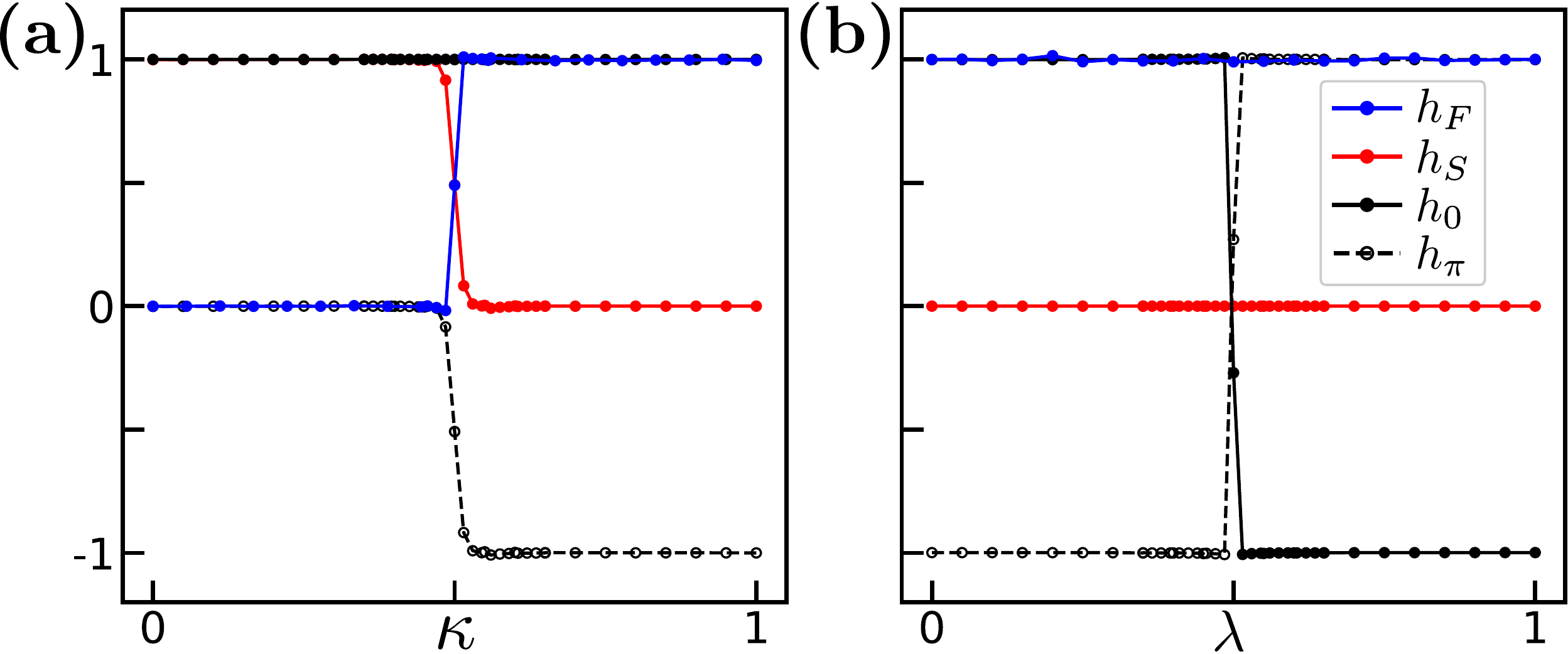}
\caption{
Numerical calculation of the Floquet invariant, static invariant, and $0$/$\pi$-defect charges \textbf{(a)} across a phase transition $(h_S, h_F) = (1, 0) \rightarrow (0,1)$ ($\kappa = 0 \rightarrow 1$) \textbf{(b)} along the smooth deformation [Eq.~(\ref{deformation})] exchanging defect charge.
} 
\label{fig: numerics}
\end{figure}
}


\begin{abstract}

We predict the existence of a novel Floquet topological insulator in three-dimensional two-band systems, the Floquet Hopf insulator, which possesses two \emph{distinct} topological invariants. One is the Hopf $\mathbb{Z}$ invariant, a linking number characterizing the (non-driven) Hopf topological insulator. The second invariant is an intrinsically Floquet $\mathbb{Z}_2$ invariant, and represents a condensed matter realization of the topology underlying the Witten anomaly in particle physics. Both invariants arise from topological defects in the system's time-evolution, subject to a process in which defects at different quasienergy exchange even amounts of topological charge. Their contrasting classifications lead to a measurable physical consequence, namely, an unusual bulk-boundary correspondence where gapless edge modes are topologically protected, but may exist at \emph{either} 0- or $\pi$-quasienergy. Our results represent a phase of matter beyond the conventional classification of Floquet topological insulators.


\end{abstract}

\maketitle

%
%

Periodically driven systems host a rich variety of phases of matter, many of which cannot be realized by any static Hamiltonian~\cite{moessner2017equilibration,zhang2017observation,choi2017observation,titum2016anomalous,maczewsky2017observation,khemani2016phase,von2016phase,kolodrubetz2018topological,li2018realistic}. 
Prime representatives of this are the so-called Floquet topological insulators (FTIs): non-interacting, driven phases of matter, whose physical properties are characterized by a set of underlying quantized topological invariants~\cite{Kitagawa2010,Gu_2011,Lindner_2011,Rudner2012,rechtsman2013photonic,Asboth2014,Carpentier2014,Nathan2015,Fruchart2016,Roy2017}. 
Unlike their non-driven counterparts, the topology of FTIs arises directly from the unitary time-evolution, leading to robustly protected gapless edge modes even when the stroboscopic time-evolution is topologically trivial. 

A common pattern has emerged in the classification of Floquet topological insulators, which relates their topological invariants to those of static topological insulators with the same dimension and symmetries. A given FTI is found to possess all the invariants of its static counterpart, plus one additional invariant of \emph{identical} classification. Intuitively, this is understood by extending the bulk-boundary correspondence to Floquet systems: under periodic modulation, the energy -- now, quasienergy -- becomes defined only modulo $2\pi$ (in units of the driving frequency) and thus an additional and identically classified edge mode emerges, associated with the bulk gap at quasienergy $\pi$.  

This result has been established rigorously in systems described by K-theory~\cite{Roy2017}, and explored at great length in the context of specific symmetries and dimensionality~\cite{Kitagawa2010,Rudner2012,Asboth2014,Carpentier2014,Nathan2015,Fruchart2016}. Nevertheless, one could wonder whether these arguments leave room for more unique topology in Floquet phases that escape this stringent bulk-boundary correspondence.

\FigureDecomposition

\FigureHomotopy

In this Letter we answer the above inquiry in the affirmative, demonstrating a three-dimensional Floquet topological insulator characterized by two \emph{distinct} topological invariants: a `static' $\mathbb{Z}$ invariant, and a uniquely Floquet $\mathbb{Z}_2$ invariant. 
At the heart of our proposal is the Hopf insulator (HI)~\cite{moore_topological_2008,deng_hopf_2013,deng_systematic_2014,deng2018probe,Kennedy2016,liu_symmetry_2017,yuan2017observation,schuster2019realizing}, a 3D topological insulator (TI) in the absence of symmetries, which exists beyond the standard K-theoretic classification~\cite{Kitaev2009,Schnyder2009} via its restriction to two-band systems. 
The $\mathbb{Z}$ invariant of our system is precisely the Hopf invariant of this insulator. 
The $\mathbb{Z}_2$ invariant replaces the expected additional integer invariant, and characterizes the same topology that underlies the Witten anomaly in (3+1)D $SU(2)$ gauge theories~\cite{Witten1982,Witten1983,Nair1984,Baer1999,Fukui2008}.
In our context, it can be understood both as a twisting number extension of the Hopf invariant, as well as in terms of gapless topological defects of the Floquet evolution. 
These `Hopf' defects may smoothly exchange even amounts of their topological charge, which leads to the reduced $\mathbb{Z}_2$ classification. 
Physically, the difference in invariants creates an atypical bulk-boundary correspondence, where gapless edge modes are topologically protected but may occur at either $0$- or $\pi$-quasienergy, depending on non-universal properties of the boundary.

%
%

We are concerned with non-interacting systems governed by a space- and time-periodic Hamiltonian, written in momentum-space as $H(\b{k},t) = H(\b{k},t+T)$, where $H(\b{k},t)$ is a matrix acting on the internal degrees of freedom that form the two bands of the system. Time-evolution is captured by the unitary operator $U(\b{k},t) = \mathcal{T} \big( e^{-i \int_0^t H(\b{k},t') dt'} \big), \, 0 \leq t < T$. Much like static insulators, one can view these unitaries in terms of the band-structures composed by their eigenvectors and eigenphases. For a two-band unitary we write
\begin{equation}\label{eigenvectors}
\begin{split}
U(\b{k},t) = e^{i \phi}\dyad{z} + e^{i \phi^\prime}\dyad{z^\prime},
\end{split}
\end{equation}
where $\phi^{(\prime)}(\b{k},t)$, $\ket{z^{(\prime)}(\b{k},t)}$ depend on time as well as momentum, and the quasienergies $\phi^{(\prime)}(\b{k},t)$ are periodic.

Floquet topological insulators are Floquet-Bloch systems where the unitary is gapped \emph{at time} $T$. The unitary $U(\b{k},T)$ is titled the Floquet unitary, and is equivalently described by the fictitious, time-independent Floquet Hamiltonian, $H_F(\b{k}) = -i \log(U(\b{k},T)) / T$. Similar to static TIs, two FTIs are in the same phase if one can smoothly interpolate between them without closing the gaps of the Floquet unitary. Much recent work has focused on characterizing these systems, revealing anomalous phases that arise from the topology of the full evolution $U(\b{k},t)$, and not just the Floquet Hamiltonian. Nevertheless, some common classifications have emerged. Focusing on Floquet unitaries with two band gaps for simplicity, in all situations with static analogues (i.e. in the absence of explicitly Floquet symmetries, e.g. time-glide symmetry~\cite{Morimoto2017}), the FTI is labelled by \emph{two} topological invariants, each with the same classification as the static TI of the same dimension and symmetries~\cite{Kitagawa2010,Rudner2012,Asboth2014,Carpentier2014,Nathan2015,Fruchart2016,Roy2017,Morimoto2017}. In Ref.~\cite{Nathan2015} these invariants are associated with band gaps of the Floquet unitary, and correspond to topological defects, or ``singularities'', in the Floquet evolution. The defects are found to be classified by the static topological invariant in a range of systems, establishing the above scheme. Ref.~\cite{Roy2017} defines the invariants differently, yet finds a similar classification for all unrestricted-band systems (where classification by K-theory is appropriate).

Here we find a finer distinction in the classification of FTIs with fixed band number. We decompose the unitary evolution into two components: the evolution over a full period, captured by the Floquet unitary $U(\b{k},T)$, and that within a period, captured by the micromotion unitary, $U_m(\b{k},t) \equiv U(\b{k},t) \, \big[ U(\b{k},T) \big]^{-t/T}$. From this decomposition, one sees that the classification factorizes into two potentially distinct invariants: a `static' invariant classifying the Floquet Hamiltonian $H_F(\b{k})$, and an intrisically Floquet invariant classifying the micromotion operator $U_m(\b{k},t)$~\cite{fn3}. In $d$ space dimensions, the former classifies maps from the $d$D Brillouin zone to the set of gapped Hamiltonians, identical to the scheme for static TIs. The Floquet invariant classifies maps from the ($d$+1)D Floquet Brillouin zone, parameterized by $(\b{k},t)$, to $SU(n)$, for an $n$-band system without symmetries~\cite{Rudner2012}. These invariants are identical in all cases previously considered. However, for systems with fixed band number they may differ.

We now introduce the Floquet Hopf insulator, a three-dimensional Floquet-Bloch system with two bands and no symmetries. The static invariant is the Hopf invariant of the Floquet Hamiltonian, which we briefly review. The gapped two-band Hamiltonian 
\begin{equation}\label{HF}
H_F(\b{k}) = \b{n}(\b{k}) \cdot \boldsymbol{\sigma}
\end{equation}
maps the 3D Brillouin zone to the Bloch sphere $S^2$. Neglecting ``weak'' lower-dimensional invariants~\cite{moore_topological_2008,Kennedy2015}, such maps are classified by the homotopy group $\pi_3(S^2) = \mathbb{Z}$, thus possessing an integer topological invariant -- the Hopf invariant. Expressed in terms of the eigenvectors $\ket{z(\b{k},T)}$ of the Floquet unitary, it takes the form
\begin{equation}\label{h_S}
h_S =  \frac{1}{2} \int d^3\b{k} \, \epsilon^{ijk} \, \mathcal{A}_i \, \mathcal{F}_{jk},
\end{equation}
where we define the Berry connection $\mathcal{A}_i = \frac{-i}{4\pi} ( \braket{z}{\partial_i z} - \braket{\partial_i  z}{z}  )$ and curvature $\mathcal{F}_{jk} = \frac{-i}{4\pi} (\braket{\partial_j z}{\partial_k z}-\braket{\partial_k z}{\partial_j z})$, and $\ket{z(\b{k},T)}$ is related to $\b{n}(\b{k})$ on the Bloch sphere by $\b{n}(\b{k}) = \bra{z(\b{k},T)} \boldsymbol{\sigma} \ket{z(\b{k},T)}$. The prototypical model HI is defined by
\begin{equation}\label{z hopf}
z(\b{k},T) \sim 
\begin{pmatrix}
\sin(k_x) + i \sin(k_y) \\
\sin(k_z) + i \big[ \sum_\alpha \cos(k_\alpha)  - m \big] \\
\end{pmatrix},
\end{equation}
which gives Hopf invariant $h_S = 1$ for $1 < m < 3$, and $h_S = 0$ for $| m | > 3$~\cite{moore_topological_2008,deng_hopf_2013}.

The Hopf invariant has an intriguing visual interpretation as a linking number. To elaborate, consider the pre-image of any $\b{n}'$ on the Bloch sphere, i.e. the set of all $\b{k}$ that are mapped to $\b{n}'$ by $\b{n}(\b{k})$. This is generically a 1D loop in the Brillouin zone. The topology of the Hopf insulator enters when one considers two such pre-images. In the HI phase, any two pre-images are \emph{linked}, with a linking number equal to the Hopf invariant. Intriguingly, this linking can be equivalently viewed as a \emph{twisting} of the Jacobian of $\b{n}(\b{k})$ along a single pre-image~\cite{Pontrjagin2007,Kennedy2016} (see Fig.~\ref{fig: decomposition}).

We now turn to the Floquet invariant. The micromotion operator maps the 4D Floquet Brillouin zone~\cite{fn1} to $SU(2)$, isomorphic to the 3-sphere $S^3$. Again neglecting weak invariants, this is classified by the group $\pi_4(S^3) = \mathbb{Z}_2$: a parity invariant, different from the integer Hopf invariant! This invariant was previously studied as the foundation of the Witten anomaly in $SU(2)$ gauge theories, where a formula for it was introduced~\cite{Witten1983}. In terms of the micromotion operator's eigenvectors $\ket{z_m(\b{k},t)}$ and their relative eigenphase $\Delta \phi_m(\b{k},t)$, we find~\cite{suppinfo}
\begin{equation}\label{h_F}
h_F =  \frac{1}{4\pi} \int dt \, d^3\b{k} \, \epsilon^{ijkl} \, \partial_i \Delta \phi_m(\b{k},t) \, \mathcal{A}_j \, \mathcal{F}_{kl} \text{ mod 2}, \\
\end{equation}
where the Berry connection and curvature are defined analogous to those in Eq.~(\ref{h_S}), now over space-time indices $\{ k_x,k_y,k_z,t \}$. The Floquet invariant also relates to the Jacobian twisting along a 1D pre-image, now in (3+1)D (see Fig.~\ref{fig: decomposition}). The higher dimensionality leads to the reduced $\mathbb{Z}_2$ classification~\cite{Pontrjagin2007, suppinfo}, familiar from the `belt trick' argument in spin-statistics~\cite{kauffman2001knots}.

\FigureNumerics

Combining the two invariants, we conclude that the Floquet Hopf insulator has a $\mathbb{Z} \times \mathbb{Z}_2$ classification. A system with arbitrary $(h_S,h_F)$ can be generated by strobing two flat band Hamiltonians according to
\begin{equation}
H_{(h_S,h_F)}(\b{k},t) = 
\begin{cases} 
\frac{2\pi}{T} H_{h_S - h_F}(\b{k}) & 0 \leq t < T/2  \\
-\frac{\pi}{T} H_{h_S}(\b{k}) & T/2 \leq t < T  \\ 
\end{cases},
\end{equation}
where $H_{h}(\b{k})$ has Hopf invariant $h$ and energies $\pm 1$. To verify the static invariant, note that the total effect of the first evolution is a minus sign, $e^{ -i \pi H} = -\mathbbm{1}$. The Floquet unitary is therefore given by $U(\b{k},T) = - e^{ i \frac{\pi}{2} H_{h_S}}$, whose bands correctly have Hopf invariant $h_S$. The Floquet invariant is also verified~\cite{suppinfo}: schematically, the contributions of the two halves of the evolution subtract, giving Floquet invariant $h_S - (h_S - h_F) = h_F \text{ mod } 2$. 

%
%

It is illuminating to discuss how the Floquet Hopf insulator fits in the context of Ref.~\cite{Nathan2015}. Here one again views the evolution in terms of bands [see e.g. Fig.~\ref{fig: homotopy}(b)], with particular attention to fixed time-slices. If the unitary $U(\b{k},t)$ is gapped at time $t$, one may define an instantaneous \emph{static} topological invariant $\mathcal{C}(t)$ from its bands, exactly as one defines the static invariant of the Floquet unitary at time $t = T$. This invariant must be constant throughout each gapped region of the evolution, and can only change at times containing gapless points. Such points are \emph{topological defects} of the evolution, at which the eigenvectors are degenerate and $\mathcal{C}(t)$ is not defined. They possess a defect charge, equal to the total change in $\mathcal{C}(t)$ across the defect. Further, they come in two varieties, $0$-defects and $\pi$-defects, labelled by the quasienergy at which the gap closes (assuming $\phi = - \phi'$ without loss of generality). The total charges of the $0$- and $\pi$-defects are locally conserved, and are thereby identified as the topological invariants of the evolution.  As an example, take the Floquet Chern insulator~\cite{Rudner2012}. The instantaneous Chern number changes at gapless Weyl points ~\cite{wan2011topological}, and the integer charges of the $0$- and $\pi$-Weyl points comprise a $\mathbb{Z} \times \mathbb{Z}$ classification~\cite{Nathan2015}.

Like other topological defects, Hopf topological defects possess an integer charge $h_{0/\pi}$ equal to the change in the instantaneous Hopf invariant across the defect. Two types of Hopf defect exist, each depicted in Fig.~\ref{fig: homotopy}(a). The first occurs a single gapless point with a quadratic energy degeneracy, exemplified by Eq.~(\ref{z hopf}) at $m = 1, \, \b{k} = 0$. Interestingly, the Hopf invariant may also change across \emph{loops} of gapless points. These arise e.g. when adding an on-site splitting $\mu \sigma_z$ to Eq.~(\ref{HF}), and may occur within a single time-slice or over a range of times. They have linear energy degeneracy and feature a Weyl cone~\cite{wan2011topological} at each point, with the frame of the Weyl cone rotating by $2\pi \Delta h$ about the loop~\cite{liu_symmetry_2017,suppinfo}.

How does conservation of the two integer defect charges reconcile with the correct $\mathbb{Z} \times \mathbb{Z}_2$ classification? The answer lies in a smooth deformation that exchanges even charge between the $0$- and $\pi$-defects, such that $(h_0,h_\pi) \rightarrow (h_0 - 2 , h_\pi + 2)$. This process has no analogue in previously studied FTIs and keeps both band gaps of the Floquet unitary open, establishing the two configurations as being of the same phase. The total charge $h_0 + h_\pi$ is conserved in this process, while the individual charge $h_\pi$ is only conserved mod 2. This suggests the identifications
\begin{equation}\label{defect_to_invariant}
\begin{alignedat}{2}
h_S & = h_0 + h_\pi && \in\,  \mathbb{Z} \\
h_F & = h_\pi \text{ mod } 2 && \in \, \mathbb{Z}_2. \\
\end{alignedat}
\end{equation}
The former follows from the definition of defect charge: the invariant at $t=T$ equals the sum of all changes to it throughout the evolution. The Floquet invariant is verified in the Supplemental Material~\cite{suppinfo}.

We explicitly describe the above deformation for the specific case of $(h_0, h_\pi) = (1,-1) \rightarrow (-1,1)$, finding a continuous family of evolutions $U(\b{k},t;\lambda)$ with defect charges $(1,-1)$ at $\lambda = 0$ and  $(-1,1)$ at $\lambda=1$ [see Fig.~\ref{fig: homotopy}(b)]. Note that the static Hopf invariant, $h_S$, vanishes in each configuration. Recall that $SU(2)$ is topologically equivalent to the 3-sphere via the parameterization
\begin{equation}
U(\b{k},t) = \xi(\b{k},t) \, \mathbbm{1} + i \b{n}(\b{k},t) \cdot \boldsymbol{\sigma},
\end{equation}
with components $(n_x,n_y,n_z,\xi) \! \in \! \mathbb{R}^4$ obeying $\xi^2 + \b{n}(\b{k})^2 = 1$. The deformation acts as a time-dependent rotation of $U(\b{k},t)$ in the $\xi n_z $-plane:
\begin{equation}\label{deformation}
U(\b{k},t;\lambda) = R_{\xi n_z}[\lambda \theta(t)] \{ U(\b{k},t) \},
\end{equation}
where the rotation angle $\lambda \theta(t)$ interpolates from $0$ at $t = 0$ to $\lambda \pi$ at times after the earliest defect (thus preserving the boundary condition $U(\b{k},t=0) = \mathbbm{1}$).

To observe that this interpolates between the two configurations without closing the Floquet gap, we examine five regions of the $\lambda=0$ and $\lambda=1$ evolutions, shown in Fig.~\ref{fig: homotopy}(c). First we consider the early and late gapped regions of the evolution, whose trivial topology allows their images to be contracted to a point. Throughout the rotation they remain points, and so stay gapped with trivial Hopf invariant. Next, we consider the $0$- and $\pi$- defects. Note that $\xi$ is related to the quasienergy as $\xi = \cos(\phi)$, so $0$-defects correspond to the North pole of the 3-sphere ($\xi = 1$) and $\pi$-defects to the South ($\xi = -1$). Upon a $\pi$-rotation in the $\xi n_z$-plane the two are interchanged, so the defects' ordering at $\lambda=0$ is reversed at $\lambda=1$.

To verify that the defect charges $h_0,h_\pi$ have changed sign at the end of the deformation, we turn to the middle region, gapped with an instantaneous Hopf invariant $h(T/2) = 1$ at $\lambda=0$. Since the times of the $0$- and $\pi$-defects are interchanged by the deformation, this time-slice must have the \emph{same} invariant $h(T/2) = 1$ at $\lambda=1$ in order to reverse the sign of the defect charges. To show that this occurs, note that the effect of the rotation on the initial eigenvectors, described by $\b{n}(\b{k},T/2)$, is to take $n_z \rightarrow -n_z$ and leave $n_x,n_y$ unchanged. These are topologically equivalent to the `flipped' eigenvectors $-\b{n} = (-n_{x},-n_{y},-n_{z})$, as the two differ by a continuous rotation about the $n_z$-axis. The Hopf invariant of $-\b{n}$ is identical to that of the initial eigenvectors $\b{n}$~\cite{fn2}, completing our verification of the deformation.

Critically, this relation does \emph{not} hold for TIs described by K-theory (e.g. the Chern insulator). Here the invariant is additive between bands, and the requirement that a fully filled system have trivial topology implies that the bands' invariants sum to zero. For two bands, this implies $\b{n}$ and $-\b{n}$ give opposite invariants. The deformation no longer does anything surprising: although it interchanges the location of the $0$- and $\pi$-defects, the invariant of the intermediate region also changes sign, and so the defect charges remain unchanged.

What allows the seemingly-conserved defect charges to change? Recall how defect charge is rigorously defined: one encloses the defect with a surface of gapped points, and computes the static topological invariant of the surface's eigenvectors~\cite{Nathan2015}. As shown in Fig.~\ref{fig: homotopy}(d), during the deformation the defects become loops of gapless points. At some value of $\lambda$, the $0$- and $\pi$-defect loops \emph{link} such that it is impossible to separately enclose each defect, causing the individual defect charges to be undefined. This defect linking arises directly from the linking of the HI~\cite{suppinfo}. After linking, the defects again have well-defined charges, which may differ from their initial values.


We compute the invariants and defect charges numerically in two scenarios  (see Fig.~\ref{fig: numerics}). Across a phase transition $(h_S, h_F) = (1, 0) \rightarrow  (0, 1)$, both the invariants and defect charges change, following Eq.~(\ref{defect_to_invariant}). In contrast, along the deformation Eq.~(\ref{deformation}), the invariants remain robustly quantized while the defects exchange charge~\cite{suppinfo}.
%
%

Like its static counterpart~\cite{moore_topological_2008,deng_hopf_2013,schuster2019realizing}, the Floquet Hopf insulator features gapless edge modes at smooth boundaries between phases with different topological invariants~\cite{fn4}. An unusual situation occurs at boundaries where the static invariant changes, but the defect charge parities do not. Here, a gap closing is protected by the change in invariant, but may occur at either $0$- \emph{or} $\pi$-quasienergy, depending on details of the edge region. The anomalous $\mathbb{Z} \times \mathbb{Z}_2$ classification is precisely what allows this ambiguity: since the defect charges are only defined up to parity, neither quasienergy individually requires a gap closing, despite the change in topological invariant.

%
%

We hope our work opens the door to additional topological phases that escape the typical correspondence between driven and non-driven classifications. It would also be interesting to search for other phases characterized by the $\mathbb{Z}_2$ twisting invariant, and to further elucidate its physical signatures. Analogues of the Floquet Hopf insulator with a stabilizing crystalline symmetry~\cite{liu_symmetry_2017}, and in the many-body localized~\cite{ponte2015periodically,khemani2016phase,von2016phase,kolodrubetz2018topological} or prethermal~\cite{abanin2015exponentially,kuwahara2016floquet,else2017prethermal} regimes, where Floquet phases can remain robust to interactions, are open problems.

\emph{Acknowledgments}: This work was supported by the DARPA DRINQS program, the Packard foundation and NIST. T.S. acknowledges support from the National Science Foundation Graduate Research Fellowship Program under Grant No. DGE 1752814.
SG acknowledges support from the Israel Science Foundation,
Grant No. 1686/18.

\bibliographystyle{apsrev4-1}
\bibliography{refs_floquet_hopf} 

\begin{thebibliography}{48}
\bibitem [{fn1()}]{fn1}%
  \BibitemOpen
  \href@noop {} {\bibinfo  {journal} {More precisely, as the micromotion
  operator must be the identity at $t = 0 \, \text{ mod } \, T$, the domain is
  the 4-torus $T^{4}$ with the sets $\{(\b{k},t) | t = 0\}$, $\{(\b{k},t) | t =
  T\}$ identified to points. In algebraic topology this is known as the
  suspension of the $3$-torus, $S T^3$}\ }\BibitemShut {NoStop}%
\bibitem [{fn2()}]{fn2}%
  \BibitemOpen
  \href@noop {} {\bibinfo  {journal} {This follows from Eq.~(3), since
  $\mathcal{F}_{jk}$ and $\mathcal{A}_i$ are each third-order in $\hat{\b{n}}$,
  making the Hopf invariant sixth-order, and thus even, in $\hat{\b{n}}$. The
  former follow from $\mathcal{F}_{jk} = \frac{1}{8\pi} \hat{\b{n}} \cdot
  (\partial_j \hat{\b{n}} \times \partial_k \hat{\b{n}})$ and $\partial_j
  \mathcal{A}_k - \partial_k \mathcal{A}_j = \mathcal{F}_{jk}$}\ }\BibitemShut
  {NoStop}%
\end{thebibliography}

\clearpage 

\section*{Supplementary material}

# Supplemental Material: Floquet Hopf Insulator

Thomas Schuster,[1] Snir Gazit,[2,1] Joel E. Moore,[1,3] and Norman Y. Yao[1,3]

[1]*Department of Physics, University of California, Berkeley, California 94720 USA*
[2]*Racah Institute of Physics and the Fritz Haber Center for Molecular Dynamics,*
*The Hebrew University, Jerusalem 91904, Israel*
[3]*Materials Science Division, Lawrence Berkeley National Laboratory, Berkeley, California 94720, USA*

(Dated: March 6, 2019)

## CONTENTS

## TWISTING INTERPRETATION OF THE TOPOLOGICAL INVARIANTS

We elaborate on the twisting interpretations of the static and Floquet Hopf invariants. Formally, these derive from the framed-cobordism understanding of the homotopy groups $\pi_3(S^2)$ and $\pi_4(S^3)$ in mathematics [1]. For the static Hopf invariant, this understanding was related to the conventional linking interpretation by Ref. [2], which we briefly summarize. Consider two infinitesimally close points $\mathbf{n}$ and $\mathbf{n} + \delta\mathbf{n}$ on the Bloch sphere. Their pre-images will also be infinitesimally close, and we may parameterize them by the periodic parameter $\theta$, $\mathbf{k}(\theta)$ and $\mathbf{k}(\theta) + \delta\mathbf{k}(\theta)$ [where $\theta$ interpolates between 0 and $2\pi$ along the pre-image]. By calculus, the difference $\delta\mathbf{k}(\theta)$ between the two pre-images is determined by the Jacobian $J_{ij}(\theta) \equiv \partial_{k_i} n_j(\mathbf{k})|_\theta$ along the first pre-image, according to $J(\theta)\delta\mathbf{k}(\theta) = \delta\mathbf{n}$. Crucially, this implies that all information about the pre-images' linking, and thus the Hopf invariant, can be found from the Jacobian along a single pre-image! The Jacobian is a $2 \times 3$ matrix (as $\mathbf{n}$ is normalized, only derivatives with respect to its two orthogonal components are nonzero), which generically has two independent non-parallel column vectors. As shown in Fig. 1 of the main text, we may choose one of these vectors to point along the direction to the second pre-image, such that the vector 'traces out' the second pre-image as it varies along the first pre-image. As seen in the figure, the total rotation, or *twisting*, of the column vectors along the pre-image is precisely the linking number of the pre-images. To formalize this, we view the two vectors as those of a real orthogonal matrix in $SO(2)$ (the reduction of the $2 \times 3$ matrix $J_{ij}(\theta)$ to a $2 \times 2$ matrix is proper, because the derivative $\partial_{k_\parallel} n_j(\mathbf{k})|_\theta$ along the pre-image is always zero). The twisting of this matrix along the 1-dimensional pre-image is classified by the homotopy group $\pi_1(SO(2)) = \mathbb{Z}$, agreeing with the original $\pi_3(S^2)$ classification.

This twisting interpretation generalizes to the Floquet invariant much better than the linking number interpretation. Although the pre-images of $SU(2)$ in the 4D Floquet Brillouin zone are again 1D loops (because $SU(2)$ is three-dimensional), there is no concept of linking number in 4D, as any two loops may be smoothly unlinked by moving in the fourth dimension. Nevertheless, one may consider the Jacobian along the pre-image loop, now a $3 \times 4$ matrix with three independent non-parallel column vectors. To analyze the twisting of the Jacobian along the loop, we view these three vectors as the columns of a real orthogonal matrix in $SO(3)$. The twisting of this matrix along the 1D loop is characterized by the group $\pi_1(SO(3)) = \mathbb{Z}_2$, reproducing the Floquet invariant's classification. The reduced classification indicates that any even twisting of the Jacobian may be smoothly deformed to a constant

Jacobian, and is familiar from characterizing particle statistics (bosons vs. fermions) in three-dimensions. We see that the reduced $\mathbb{Z}_2$ invariant is the natural result of extending the static Hopf invariant to higher dimensions.

We can verify the twisting interpretation for the stroboscopic evolution described in the main text. To do so, we first write down the micromotion operator for the stroboscopic evolution. The micromotion operator closes the bands of the Floquet unitary, and is homotopy equivalent to the flat band unitary:

$$U_m(\mathbf{k}, t) = \begin{cases} e^{-\frac{2\pi i t}{T} H_{h_S - h_F}(\mathbf{k})} & 0 \leq t < T/2 \\ -e^{-\frac{2\pi i (t - T/2)}{T} H_{h_S}(\mathbf{k})} & T/2 \leq t < T \end{cases}. \tag{1}$$

We'll take $h_F = 1 \bmod 2$, $h_S = 0$, and choose the trivial insulator to be the product state, $H_{h_S=0}(\mathbf{k}) = \sigma_z$. Consider the pre-image of fixed $\hat{\mathbf{n}} \neq \hat{\mathbf{z}}$ and $\phi$. This occurs at fixed time-slice $t = \phi T/2\pi$, with $\mathbf{k}$ determined by the pre-image of $\hat{\mathbf{n}}$ under the static Hamiltonian $H_{h_S - h_F = 1}(\mathbf{k})$. Compared to the static case, the Jacobian along a pre-image has one new row, corresponding to derivatives of the quasienergy $\phi$, and one new column, for derivatives with respect to time $t$. In the flat band micromotion operator $\partial_k \phi = 0$ and $\partial_t \hat{\mathbf{n}} = 0$, and so the time column is nonzero only on the quasienergy row and is orthogonal to both other columns. The Jacobian twisting is therefore determined entirely by the columns corresponding to $\hat{\mathbf{n}}$, and so is the same as the twisting for the static Hamiltonian $H_{h_S - h_F}(\mathbf{k}) = H_1(\mathbf{k})$: equal to the Floquet invariant $h_F = 1$.

## DERIVATION OF THE FLOQUET INVARIANT

We compute the Floquet Hopf invariant, which characterizes elements of the homotopy group $\pi_4(SU(2)) = \mathbb{Z}_2$, by adapting a method introduced by Witten in the context of $SU(2)$ gauge theories [3]. This method relies on the fact that the fourth homotopy group of three-band unitaries, $\pi_4(U(3))$, is trivial, in contrast to the homotopy group of interest, of $SU(2)$ [or, equivalently, $U(2)$]. The prescription is as follows. First, embed the micromotion unitary $U_m(\mathbf{k}, t)$ in $U(3)$ by adding a third, trivial band (i.e. with no $\mathbf{k}, t$ dependence):

$$\tilde{U}_m(\mathbf{k}, t) \equiv \begin{pmatrix} U_m(\mathbf{k}, t) & 0 \\ & 0 \\ 0 & 0 & 1 \end{pmatrix}. \tag{2}$$

Next, contract the three-band unitary $\tilde{U}_m(\mathbf{k}, t)$ smoothly to the identity. Specifically, find a continuous family of unitaries $\tilde{U}(\mathbf{k}, t, \rho)$, labelled by contraction parameter $\rho$, such that $\tilde{U}(\mathbf{k}, t, \rho = 0) = \tilde{U}_m(\mathbf{k}, t)$ and $\tilde{U}(\mathbf{k}, t, \rho = \pi/2) = \mathbb{1}$, $\forall \mathbf{k}, t$. Prior to embedding in $U(3)$, such a contraction is forbidden if the Hopf defect at $\pi$-quasienergy has a nontrivial topological charge. After embedding, the contraction becomes possible due to the triviality of the Hopf insulator in three-band systems [or, equivalently, due to the trivial homotopy group $\pi_4(U(3))$]. Remarkably, the Floquet Hopf invariant of the *original* two-band micromotion operator can be computed from the contraction $\tilde{U}(\mathbf{k}, t, \rho)$ via the five-dimensional integral

$$h_F = \frac{-i}{240\pi^3} \int_0^{\pi/2} d\rho \int dt \, d^3\mathbf{k} \sum_{\substack{ijklm \\ k_x, k_y, k_z, t, \rho}} \epsilon^{ijklm} \text{Tr}\big(\tilde{U}_m^{-1} \partial_i \tilde{U} \tilde{U}_m^{-1} \partial_j \tilde{U} \tilde{U}^{-1} \partial_k \tilde{U} \tilde{U}^{-1} \partial_l \tilde{U} \tilde{U}^{-1} \partial_m \tilde{U}\big). \tag{3}$$

The invariant $h_F$ is quantized to an integer modulo 2, and is independent of the deformation chosen. The modding can be partially explained as an ambiguity in how one contracts $\tilde{U}(\mathbf{k}, t, \rho)$ [3].

Here, we recast the above formula in terms of the eigenvectors $|z_m\rangle$ and quasienergies $\phi_m$ of the *original* micromotion operator $U_m(\mathbf{k}, t)$. We do this by writing down an explicit contraction (valid for an arbitrary micromotion operator), and using it to simplify Eq. (3). The embedding and contraction are depicted in Fig. 1. The embedded micromotion operator has three bands: an upper band with eigenvector $\big(z_{m,1}(\mathbf{k}, t), z_{m,2}(\mathbf{k}, t), 0\big)^T$, a lower band with orthogonal eigenvector $\big(-z_{m,2}^*(\mathbf{k}, t), z_{m,1}^*(\mathbf{k}, t), 0\big)^T$, and the third, trivial band with eigenvector $\big(0, 0, 1\big)^T$. Here, the upper and lower bands are written in terms of the original two-band micromotion operator's upper eigenvector $|z_m(\mathbf{k}, t)\rangle = \big(z_{m,1}(\mathbf{k}, t), z_{m,2}(\mathbf{k}, t)\big)^T$, and are labelled in Fig. 1 by the original Bloch spin vectors $+\hat{\mathbf{n}}(\mathbf{k}, t)$, $-\hat{\mathbf{n}}(\mathbf{k}, t)$. The contraction consists of three steps. First, the quasienergy of the trivial band is shifted to be degenerate with the lower $SU(2)$ band. At this point the unitary takes the form

$$\tilde{U}(\mathbf{k}, t) = e^{i\phi_m(\mathbf{k}, t)} |\psi(\mathbf{k}, t)\rangle\langle\psi(\mathbf{k}, t)| + e^{-i\phi_m(\mathbf{k}, t)}(1 - |\psi(\mathbf{k}, t)\rangle\langle\psi(\mathbf{k}, t)|), \tag{4}$$

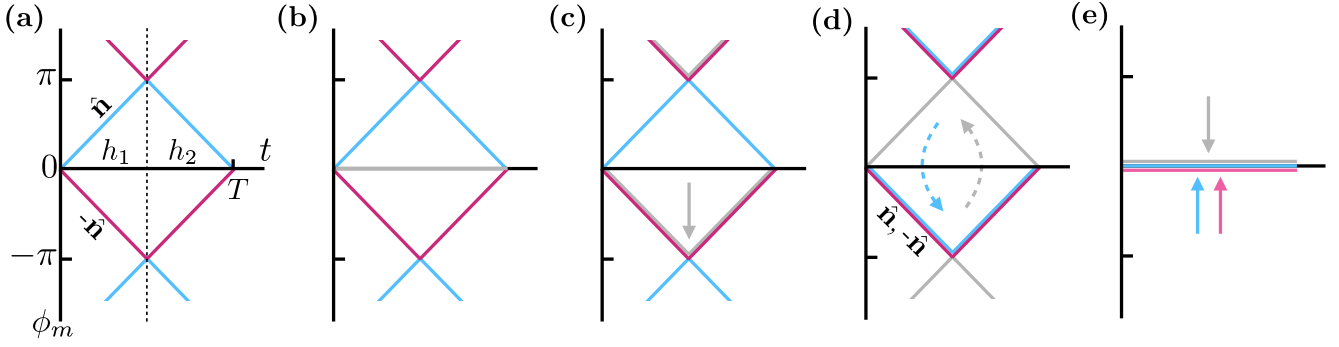

FIG. 1. Depiction of the embedding of the micromotion operator in $U(3)$, and its contraction to the identity operator. (**a**) The original, two-band micromotion operator. Its upper (blue) and lower band (magenta) are labelled by their Bloch spins $\hat{\mathbf{n}}, -\hat{\mathbf{n}}$, and possess some instantaneous Hopf invariant. Here we depict a flat band micromotion operator, where the invariant changes from $h_1$ to $h_2$ across a $\pi$-defect at $t = T/2$. If $h_1 \neq h_2$ mod 2, the $\pi$-defect is topologically protected and the unitary cannot be contracted to the identity. (**b**) After embedding, the unitary gains an additional, trivial band (gray) at zero quasienergy. (**c**) The first step of the contraction changes the quasienergy of the trivial band to be degenerate with the lower band (solid arrow). (**d**) Second, the eigenvectors of the upper band and the trivial are smoothly exchanged via Eq. (5), without changing either band's quasienergy (dashed arrows). After this exchange, the original upper and lower band are degenerate, and do not possess any nontrivial topology. (**e**) This allows the quasienergy of all bands to be deformed continuously to zero, completing the contraction.

where $|\psi(\mathbf{k}, t)\rangle = \left(z_1(\mathbf{k}, t), z_2(\mathbf{k}, t), 0\right)^T$ is the upper band's three-component eigenvector. In the second step, we use the triviality of the Hopf insulator in three-band systems to deform the upper band to a trivial band (no $\mathbf{k}$-dependence), while leaving the quasienergies $\phi_m(\mathbf{k}, t)$ unchanged. Specifically, the unitary maintains the above form, but with upper band eigenvector

$$|\psi(\mathbf{k}, t, \rho)\rangle = \begin{pmatrix} \cos(\rho) \, z_{m,1}(\mathbf{k}, t) \\ \cos(\rho) \, z_{m,2}(\mathbf{k}, t) \\ \sin(\rho) \end{pmatrix}, \tag{5}$$

for $\rho \in [0, \pi/2]$. As the lower bands' eigenvectors must remain orthogonal to the upper band, this step effectively swaps the eigenvectors of the upper band and the trivial lower band [Fig. 1(d)]. After this step, the two $\mathbf{k}$-dependent bands are degenerate (see Fig. 1), and are therefore equivalently labelled by $\mathbf{k}$-*independent* eigenvectors $(1, 0, 0)$, $(0, 1, 0)$. This relabelling makes it clear that the 'defect' at $\pi$-quasienergy is no longer topologically protected. In the third step, we complete the contraction by continuously take the quasienergies of all three bands to 0 for all $\mathbf{k}, t$.

Contraction in hand, we turn to computing the Floquet Hopf invariant via Eq. (3). The first and final steps of the contraction involve changing the quasienergies of bands with no $\mathbf{k}$-dependence, and as such can be shown to not contribute to Eq. (3). We thus only need to consider the second step, defined via Eqs. (4) and (5). Using these, we compute

$$\tilde{U}^{-1} \partial_i \tilde{U} = (i \partial_i \Delta\phi_m) \, |\psi\rangle\langle\psi| + |\alpha|^2 \, \langle\psi|\partial_i\psi\rangle \, |\psi\rangle\langle\psi| - \alpha^* \, |\psi\rangle\langle\partial_i\psi| + \alpha \, |\partial_i\psi\rangle\langle\psi| - \frac{i}{2} \partial_i \Delta\phi_m \mathbb{1}, \tag{6}$$

where we define the quasienergy difference $\Delta\phi_m \equiv 2\phi_m$, and $\alpha(\mathbf{k}, t) \equiv e^{i\Delta\phi_m(\mathbf{k}, t)} - 1$ for convenience. From this, we see that the integrand of Eq. (3) consists of a sum of terms, each of which are products of expressions of the form $i\partial_i\Delta\phi_m$, $\langle\psi|\partial_i\psi\rangle$, $\langle\partial_i\psi|\psi\rangle$, or $\langle\partial_i\psi|\partial_j\psi\rangle$, with each $\partial_i$ occurring exactly once in each term. We note that the final term $\sim \mathbb{1}$ in this expression can be eliminated by adding a global phase to $\tilde{U}(\mathbf{k}, t, \rho)$, and thus does not contribute the topological invariant.

We now argue that only terms of the form $\partial_i\phi \, \langle\partial_j\psi|\partial_k\psi\rangle \, \langle\partial_l\psi|\partial_m\psi\rangle$ are nonzero after antisymmetrization by $\epsilon^{ijklm}$. It is clear that any term with greater than one derivative acting on $\Delta\phi_m$ antisymmetrizes to zero. Similarly, any term that contains greater than one expression $\langle\psi|\partial_i\psi\rangle$ also antisymmetrizes to zero. A term containing $\langle\psi|\partial_i\psi\rangle \, \langle\partial_j\psi|\psi\rangle$ also gives zero, as $\langle\psi|\partial_i\psi\rangle$ and $\langle\psi|\partial_j\psi\rangle$ are both pure imaginary (because $|\psi\rangle$ is normalized), and so the index-flipped $\langle\psi|\partial_j\psi\rangle \, \langle\partial_i\psi|\psi\rangle = \langle\psi|\partial_i\psi\rangle^* \, \langle\partial_j\psi|\psi\rangle^*$ is equal to the original expression and they cancel after

antisymmetrization. The only terms left given these restrictions are of the form $\partial_i \phi \langle \partial_j \psi | \partial_k \psi \rangle \langle \partial_l \psi | \partial_m \psi \rangle$ and $\langle \psi | \partial_i \psi \rangle \langle \partial_j \psi | \partial_k \psi \rangle \langle \partial_l \psi | \partial_m \psi \rangle$. The latter is ruled out by our choice of deformation: recalling our expression for $|\psi\rangle$ in terms of $|z\rangle$ in Eq. (5), we compute the inner products

$$
\begin{aligned}
\langle \partial_\rho \psi | \psi \rangle &= 0 \\
\langle \partial_\rho \psi | \partial_j \psi \rangle &= -\cos(\rho)\sin(\rho)\,\langle z(\mathbf{k},t) | \partial_j z(\mathbf{k},t) \rangle \\
\langle \partial_l \psi | \partial_m \psi \rangle &= \cos^2(\rho)\,\langle \partial_l z(\mathbf{k},t) | \partial_m z(\mathbf{k},t) \rangle,
\end{aligned}
\tag{7}
$$

for $j, l, m \in \{k_x, k_y, k_z, t\}$. As $\partial_\rho \phi$ and $\langle \psi | \partial_\rho \psi \rangle$ are both zero, the $\rho$-derivative must appear in a term of the form $\partial_\rho \psi | \partial_j \psi \rangle \sim \langle z(\mathbf{k},t) | \partial_j z(\mathbf{k},t) \rangle$. However, the first expression in the term $\langle \psi | \partial_i \psi \rangle \langle \partial_j \psi | \partial_k \psi \rangle \langle \partial_l \psi | \partial_m \psi \rangle$ is similarly proportional to some $\langle z(\mathbf{k},t) | \partial_i z(\mathbf{k},t) \rangle$, and so the term will again vanish upon antisymmetrization.

Keeping only the aforementioned nonzero terms, we have

$$
\begin{aligned}
h_F &= \frac{1}{48\pi^3} \int_0^{\pi/2} d\rho \int dt\, d^3\mathbf{k}\, |\alpha|^4 \sum_{\substack{ijklm \in \\ k_x, k_y, k_z, t, \rho}} \epsilon^{ijklm} \partial_i \Delta\phi_m \langle \partial_j \psi | \partial_k \psi \rangle \langle \partial_l \psi | \partial_m \psi \rangle \\
&= \frac{1}{24\pi^3} \int d\rho\, dt\, d^3\mathbf{k}\, |\alpha|^4 \sum_{\substack{ijklm \in \\ k_x, k_y, k_z, t, \rho}} \epsilon^{ijlm} \partial_i \Delta\phi_m \big( \langle \partial_j \psi | \partial_\rho \psi \rangle - \langle \partial_\rho \psi | \partial_j \psi \rangle \big) \langle \partial_l \psi | \partial_m \psi \rangle.
\end{aligned}
\tag{8}
$$

A factor of 5 was gained from the cyclicity of the trace. We proceed further by using Eqs. (7) and performing the $\rho$ integral. We find

$$
\begin{aligned}
h_F &= -\frac{1}{24\pi^3} \left( \int d\rho \cos^3(\rho)\sin(\rho) \right) \int dt\, d^3\mathbf{k}\, |\alpha|^4 \sum_{\substack{ijlm \in \\ k_x, k_y, k_z, t}} \epsilon^{ijlm} \partial_i \Delta\phi_m \big( \langle z(\mathbf{k},t) | \partial_j z(\mathbf{k},t) \rangle - \langle \partial_j z(\mathbf{k},t) | z(\mathbf{k},t) \rangle \big) \langle \partial_l z(\mathbf{k},t) | \partial_m z(\mathbf{k},t) \rangle \\
&= -\frac{1}{96\pi^3} \int dt\, d^3\mathbf{k}\, |\alpha|^4 \sum_{\substack{ijlm \in \\ k_x, k_y, k_z, t}} \epsilon^{ijlm} \partial_i \Delta\phi_m \big( \langle z(\mathbf{k},t) | \partial_j z(\mathbf{k},t) \rangle - \langle \partial_j z(\mathbf{k},t) | z(\mathbf{k},t) \rangle \big) \langle \partial_l z(\mathbf{k},t) | \partial_m z(\mathbf{k},t) \rangle, \\
&= \frac{1}{24\pi} \int dt\, d^3\mathbf{k} \left[ 6 - 8\cos(\Delta\phi_m) - 2\cos(2\Delta\phi_m) \right] \sum_{\substack{ijlm \in \\ k_x, k_y, k_z, t}} \epsilon^{ijlm} \partial_i \Delta\phi_m\, \mathcal{A}_j\, \mathcal{F}_{kl},
\end{aligned}
\tag{9}
$$

defining $\mathcal{A}_j = \frac{-i}{4\pi}(\langle z | \partial_j z \rangle - \langle \partial_j z | z \rangle)$ and $\mathcal{F}_{lm} = \frac{-i}{2\pi}(\langle \partial_l z(\mathbf{k},t) | \partial_m z(\mathbf{k},t) \rangle - \langle \partial_m z(\mathbf{k},t) | \partial_l z(\mathbf{k},t) \rangle)$, and recalling $|\alpha|^4 = 6 - 8\cos(\Delta\phi_m) - 2\cos(2\Delta\phi_m)$. We will soon show that only the first, non-oscillating term of $|\alpha|^4$ is nonzero after integration, in which case the above expression replicates the Floquet Hopf invariant formula presented in the main text. We have also confirmed that these terms are zero numerically.

We are now ready to confirm the relation between the Floquet invariant and the defect charges Eq. (7) of the main text. We begin with the special case of flat band micromotion operators, to build intuition. Consider the micromotion operator depicted in Fig. 1(a) (and written in Eq. (1), replacing $h_S - h_F \to h_1, h_S \to h_2$). Here $\Delta\phi_m(\mathbf{k},t) = \Delta\phi_m(t)$ and so we have

$$
\begin{aligned}
h_F &= \frac{1}{24\pi} \left( \int_0^T dt\, \partial_t \Delta\phi_m \left[ 6 - 8\cos(\Delta\phi_m) - 2\cos(2\Delta\phi_m) \right] \right) \left( \int d^3\mathbf{k} \sum_{\substack{jlm \in \\ k_x, k_y, k_z}} \epsilon^{jlm}\, \mathcal{A}_j\, \mathcal{F}_{kl} \right) \\
&= \frac{1}{12\pi} \int_0^T dt\, (\partial_t \Delta\phi_m) \left[ 6 - 8\cos(\Delta\phi_m) - 2\cos(2\Delta\phi_m) \right] h(t) \\
&= \frac{1}{12\pi} \left( \int_0^{2\pi} d(\Delta\phi_m) \left[ 6 - 8\cos(\Delta\phi_m) - 2\cos(2\Delta\phi_m) \right] h_1 + \int_{2\pi}^0 d(\Delta\phi_m) \left[ 6 - 8\cos(\Delta\phi_m) - 2\cos(2\Delta\phi_m) \right] h_2 \right) \\
&= \frac{6 \times 2\pi}{12\pi} \left( h_1 - h_2 \right) \\
&= h_\pi \bmod 2.
\end{aligned}
\tag{10}
$$

In the first step, we use the definition of the Hopf invariant in Eq. (3) of the main text. In the second to last step, we separately compute the integral for the two regions in which $h(t)$ is constant: $t \in [0, T/2]$, where $\Delta\phi_m$ increases from 0 to $2\pi$, and $t \in [T/2, T]$, where $\Delta\phi_m$ decreases from $2\pi$ to 0. The integral receives a positive contribution from the first region, and a negative contribution from the second; this subtraction is precisely the change in Hopf invariant between the two regions, at $\Delta\phi_m = 2\pi$ ($\phi_m = \pi$). It is straightforward to generalize this to the case of multiple defects, and one replicates the above formula for the total $\pi$-defect charge $h_\pi$.

Inspired by the flat band case, we can relate the invariant to the defect charges for general micromotion operators with the help of a coordinate change. We transform from the space-time coordinates $(k_x, k_y, k_z, t)$ to coordinates $(x_1, x_2, x_3, \Delta\phi_m)$, where $\Delta\phi_m$ is again the micromotion's relative phase and the $x_i$ are some coordinates that parameterize the three-dimensional submanifolds of constant $\Delta\phi_m$. In general this coordinate transformation is not possible (not 1-to-1) on the entire space, and we will have to divide the space into a union of patches $P$, each with their own coordinates $(x_1^P, x_2^P, x_3^P, \Delta\phi_m^P)$. This is similar to the flat band case, where $\Delta\phi_m \sim t$ is increasing on the first patch ($t < T/2$) and $\Delta\phi_m \sim -(t - T/2)$ is decreasing on the second ($t > T/2$). The Floquet Hopf invariant then takes the form

$$h_F = \frac{1}{24\pi} \sum_P \int d(\Delta\phi_m^P) \, |\alpha(\Delta\phi_m^P)|^4 \int d^3\mathbf{x}^P \sum_{jkl \in \{x_1^P, x_2^P, x_3^P\}} \epsilon^{jkl} \, \mathcal{A}_j(\Delta\phi_m^P, \mathbf{x}^P) \, \mathcal{F}_{kl}(\Delta\phi_m^P, \mathbf{x}^P), \tag{11}$$

where the change of coordinates has little effect because the integral is a topological invariant. The $\mathbf{x}^P$-integrals are performed over three-dimensional manifolds, defined by the surface of constant $\Delta\phi_m^P$. Contributions from different patches of the same constant $\Delta\phi_m$ are added together, each with the same prefactor $|\alpha(\Delta\phi_m)|^4$. The sum over $P$ thus amounts to changing the domain of the $\mathbf{x}$-integral to be over the *entire* submanifold of constant $\Delta\phi_m$ (i.e. over all patches) , which we will call $\mathcal{M}_{\Delta\phi_m}$. We have,

$$h_F = \frac{1}{12\pi} \int d(\Delta\phi_m) \, |\alpha(\Delta\phi_m)|^4 \left[ \frac{1}{2} \int_{\mathcal{M}_{\Delta\phi_m}} d^3\mathbf{x} \sum_{jkl \in \{x_1, x_2, x_3\}} \epsilon^{jkl} \, \mathcal{A}_j(\Delta\phi_m, \mathbf{x}) \, \mathcal{F}_{kl}(\Delta\phi_m, \mathbf{x}) \right]. \tag{12}$$

The submanifold $\mathcal{M}_{\Delta\phi_m}$ is closed, and may consist of multiple disjoint regions. The sign of the contribution from each region is equivalent to the orientation of that region, and is fixed by sign of the original derivative $(\partial_i \Delta\phi_m)$ to point in the direction of increasing $\Delta\phi_m$. (For instance, in the flat band example there are two regions for each $\Delta\phi_m$, the increasing region $[t < T/2]$ and the decreasing region $[t > T/2]$, and the contribution from the decreasing region is subtracted from that of the increasing region.) Looking at Eq. (13), the integral over $\mathcal{M}_{\Delta\phi_m}$ is nothing more than the (signed) sum of the static Hopf invariants of each closed region of the submanifold, $h(\mathcal{M}_{\Delta\phi_m})$. As a topological invariant, $h(\mathcal{M}_{\Delta\phi_m})$ cannot change with the relative phase $\Delta\phi_m$, and is therefore fixed to be a constant for all $\mathcal{M}_{\Delta\phi_m}$. As one might guess from the flat band example, this constant is precisely the Floquet Hopf invariant,

$$\begin{aligned} h_F &= \frac{1}{12\pi} \int_0^{2\pi} d(\Delta\phi_m) \, |\alpha(\Delta\phi_m)|^4 h(\mathcal{M}_{\Delta\phi_m}) \\ &= \frac{1}{12\pi} \int_0^{2\pi} d(\Delta\phi_m) \left[ 6 - 8\cos(\Delta\phi_m) - 2\cos(2\Delta\phi_m) \right] h(\mathcal{M}_{\Delta\phi_m}) \\ &= h(\mathcal{M}_{\Delta\phi_m}). \end{aligned} \tag{13}$$

where the constancy of $h(\mathcal{M}_{\Delta\phi_m})$ causes all oscillatory terms $\sim e^{in\Delta\phi_m(\mathbf{k},t)}$ (for nonzero $n$) to integrate to zero.

It remains only to identify the constant $h(\mathcal{M}_{\Delta\phi_m}) = h_F$ with the defect charges. Consider the submanifold $\mathcal{M}_{\Delta\phi_m}$ for $\Delta\phi_m$ infinitesimally close to $2\pi$. For a micromotion operator with $N$ point Hopf defects at $\pi$-quasienergy, of charges $\Delta h_\pi^i, i = 1, \ldots, N$, this manifold consists of $N$ 3-spheres, each enclosing one of the point Hopf defects. The Hopf invariant of each 3-sphere is exactly the charge of its enclosed defect, and each 3-sphere contributes with the same sign. The total Hopf invariant of the submanifold then gives

$$h_F = h(\mathcal{M}_{2\pi-\epsilon}) = \sum_i \Delta h_\pi^i = h_\pi. \tag{14}$$

This completes our identification of the parity of $h_\pi$ with the Floquet invariant.

## EXAMPLES OF POINT AND LOOP HOPF DEFECTS

A simple example of a point Hopf defect, of charge $\Delta h = 1$, occurs in Eq. (4) of the main text, at $m = 1$, $\mathbf{k} = 0$. Replacing $m$ with $t$, the eigenvector near the defect varies as

$$|z(\mathbf{k}, t)\rangle = \begin{pmatrix} \delta k_x + i\, \delta k_y \\ \delta k_z + i\, \delta t \end{pmatrix}. \tag{15}$$

To derive the quadratic energy degeneracy of this defect, we consider the instantaneous Hamiltonian giving rise to the defect, $H_{\text{inst}}(\mathbf{k}, t) = \mathbf{n}(\mathbf{k}, t) \cdot \boldsymbol{\sigma}$, with $\mathbf{n}(\mathbf{k}, t) = \langle z(\mathbf{k}, t) | \boldsymbol{\sigma} | z(\mathbf{k}, t) \rangle$. Note that $|z(\mathbf{k}, t)\rangle$, and therefore $\mathbf{n}(\mathbf{k}, t)$, are not normalized here, in contrast to Eq. (2) of the main text. [There, we considered a gapped Hamiltonian, which can always be deformed to a flat band Hamiltonian with normalized $\mathbf{n}(\mathbf{k}, t)$.] Their normalization gives us the quasienergy in the vicinity of the defect, $\Delta\phi(\mathbf{k}, t) = |H_{\text{inst}}(\mathbf{k}, t)| = |\langle z(\mathbf{k}, t) | \boldsymbol{\sigma} | z(\mathbf{k}, t) \rangle| = \delta\mathbf{k}^2 + \delta t^2$.

We now turn to loop Hopf defects. After demonstrating an example of such a defect, we will use the example to show: first, that the *point* Hopf defect can be viewed as a contraction of the loop defect; and second, that each point along the loop defect appears as a Weyl cone, with the frame of the Weyl cone twisting by $2\pi$ as one circles the defect. Our example loop defect is also found in Eq. (4) of the main text, this time by adding an on-site splitting $\mu\, \sigma_z$ to the associated instantaneous Hamiltonian. This splitting changes the Bloch sphere vector $\mathbf{n}(\mathbf{k}) \to \mathbf{n}(\mathbf{k}) + \mu\hat{\mathbf{z}}$. In the Hopf insulator phase, the pre-image of $\hat{\mathbf{n}}(\mathbf{k}) = -\hat{\mathbf{z}}$ is a loop in the Brillouin zone defined by $k_z = 0$, $\cos(k_x) + \cos(k_y) = m - 1$, with energy $E(\mathbf{k}) = |\mathbf{n}(\mathbf{k})| = \sqrt{||z_1(\mathbf{k})|^2 - \mu|} = \sqrt{|\sin^2(k_x) + \sin^2(k_y) - \mu|}$ along the loop. For each point along the pre-image, there is a value of $\mu$ at which the energy equals zero, and above which $n_z$ is positive. As $\mu$ is increased this change propagates along the pre-image, until all points have transitioned to being pre-images of $\hat{\mathbf{n}}(\mathbf{k}) = +\hat{\mathbf{z}}$. Envisioning $\mu$ as varying in time, this realizes a loop defect with some finite extent in the time-direction.

It is illuminating and convenient to analyze the loop defect in the limit where it occurs at small $k_x, k_y$. The smallness of the pre-image is controlled by the parameter $\epsilon = 3 - m$, where for small $\epsilon$ the pre-image occurs along the circle $k_z = 0$, $k_x^2 + k_y^2 = 2\epsilon$, with energy $\sqrt{|2\epsilon - \mu|}$ [neglecting corrections $O(k^3)$]. Since the energy is now constant along the pre-image, the defect occurs at a single value of the splitting (i.e. a single instant in time), $\mu = 2\epsilon$. Near the defect, the Bloch sphere vector varies as

$$\begin{aligned} n_x &= k_x k_z + k_y \delta \\ n_y &= -k_x \delta - k_y k_z \\ n_z &= k_x^2 + k_y^2 - k_z^2 - \delta^2 - \mu \end{aligned}, \tag{16}$$

defining $k_r^2 \equiv k_x^2 + k_y^2$, $\delta \equiv \epsilon - \frac{1}{2} k_r^2$, and $\mu = 2\epsilon + \delta\mu$. Keeping only first order terms, and defining $\delta k_r \equiv -\delta/\sqrt{2\epsilon}$ and $k_x = k_r \cos(\theta)$, $k_y = k_r \sin(\theta)$ we have

$$\begin{aligned} n_x &= \sqrt{2\epsilon}\, k_z \cos(\theta) - 2\epsilon\, \delta k_r \sin(\theta) \\ n_y &= -2\epsilon\, \delta k_r \cos(\theta) + \sqrt{2\epsilon}\, k_z \sin(\theta) \cdot \\ n_z &= -\delta\mu \end{aligned} \tag{17}$$

From this we see that the energy is indeed linearly dispersing about the loop defect. We note that in the limit $\epsilon \to 0$ the linearly-dispersing loop defect contracts to a single quadratically-dispersing point at $\mathbf{k} = 0$, showing that the point Hopf defect can be understood as a contraction of the loop defect.

At each point along the loop, the three vectors $\partial_{k_z}\mathbf{n}$, $\partial_\mu\mathbf{n}$, and $\partial_{k_r}\mathbf{n}$ define the frame of a Weyl cone extending in the directions perpendicular to the loop. To analyze the twisting of the Weyl cone about the loop, we compute these vectors

$$\begin{aligned} \partial_{k_z}\mathbf{n} &= \left( \sqrt{2\epsilon} \cos(\theta),\, -\sqrt{2\epsilon} \sin(\theta),\, 0 \right) \\ \partial_{k_r}\mathbf{n} &= \left( -2\epsilon \sin(\theta),\, -2\epsilon \cos(\theta),\, 0 \right) \\ \partial_\mu\mathbf{n} &= \left( 0,\, 0,\, -1 \right). \end{aligned} \tag{18}$$

These vectors are linearly independent and twist by $2\pi$ about the $\mu$-axis along the loop parameterized by $\theta$, verifying the Weyl cone interpretation.

## DEFECTS DURING THE DEFORMATION

Here we describe the topological defects' trajectories throughout the smooth deformation in more detail. We denote the initial ($\lambda = 0$) times of the 0- and $\pi$-defects as $t_0$ and $t_\pi$ respectively. Expressing $U_\lambda$ in the 3-sphere coordinates $(\xi, \mathbf{n})$, the gapless points at a given $\lambda$ are by definition the pre-images of $U_\lambda = (\pm 1, 0, 0, 0)$. From Eq. (9) of the main text, these are identical to the pre-images of the points $R^{-1}_{\xi n_z}(\lambda)(\pm 1, 0, 0, 0) = (\pm \cos(\pi\lambda), 0, 0, \pm \sin(\lambda))$ in the *un-rotated* evolution $U_{\lambda=0}$. For simplicity, we assume that $U_0$ has flat bands for some range in between $t_0$ and $t_\pi$, such that in this range the quasienergies linearly interpolate between their defect values of 0 and $\pi$, $\xi(\mathbf{k}, t) = \xi(t) = \cos\left(\frac{t-t_0}{t_\pi-t_0}\pi\right)$. With this assumption, the defects occur at fixed time-slices $t = \xi^{-1}\left(\pm\cos(\pi\lambda)\right) = \pm(\lambda - \frac{1}{2} \pm \frac{1}{2})(t_\pi - t_0) + t_0$. (If we relax the flat band assumption the defects will have some extent in the $t$-direction.) Further, within these time-slices, the pre-images in the Brillouin zone are precisely the pre-images of $\hat{\mathbf{n}} = \pm\hat{\mathbf{z}}$ in the instantaneous *static* Hamiltonian defined by the evolutions' eigenvectors at this time. Thus the 0-defect is a loop of gapless points moving forward in time (as $\lambda$ increases) and the $\pi$-defect is a different loop moving backwards. At the intermediate value $\lambda = 1/2$, the two defects exist in the same time-slice $t = (t_0 + t_\pi)/2$ and are *linked* in the Brillouin zone, by virtue of the linking of the $\hat{\mathbf{n}} = \pm\hat{\mathbf{z}}$ pre-images in the instantaneous Hamiltonian, which has Hopf invariant 1. At this $\lambda$, only the total charge $h_0 + h_\pi$ is well-defined. The analysis of the main text shows that as the pre-images pass each other the individual charges $h_0$ and $h_\pi$ are indeed changed from their initial values, and only the total charge and the Floquet invariant $h_\pi$ mod 2 are preserved.

## DETAILS ON NUMERICS

The evolution depicted in Fig. 2 of the main text, and those used the numerics of Fig. 3 of the main text, are calculated similarly. The evolution consists of five components:

1. For times $0 \le t < t_1$ the quasienergy linearly increases, $\phi = (t/t_1)\phi_m$. For the first half of this $0 \le t < t_1/2$ the eigenvector are in the product state $\hat{\mathbf{n}}(\mathbf{k}, t) = \hat{\mathbf{z}}$. For the second half $t_1/2 \le t < t_1$ the eigenvectors are smoothly deformed to those of the trivial insulator described by Eq. (4) of the main text at $m = 4$. This done by multiplying the first component $z_1(\mathbf{k}, t)$ of the eigenvector by constant that increases from 0 to 1 from $t_1/2$ to $t_1$, and is possible while maintaining the eigenvectors normalization because $z_2(\mathbf{k}, t) \ne 0$ for all $\mathbf{k}$ in the trivial phase.

2. For times $t_1 \le t < t_2$, the eigenvectors continue to be described by Eq. (4) of the main text, but with $m$ varying with time as $m(t) = 4 - 2(t - t_1)/(t_2 - t_1)$. During this the instantaneous Hopf invariant changes from 0 at $m = 4$ to 1 at $m = 2$ through a gap closing at quasienergy 0, corresponding to a point 0-defect with charge $+1$. We take the quasienergy to be

$$\phi = \begin{cases} (1-d) + d|z(\mathbf{k}, t)|^2 & (1-d) + d|z(\mathbf{k}, t)|^2 < \phi_m \\ \phi_m & \text{else} \end{cases} \tag{19}$$

where $d \equiv e^{-(k_x^2 + k_x^2 + (m-3)^2)/w^2}$ is 1 at the gapless point and asymptotically 0. This describes a gapless region of approximately quadratic dispersion and finite-extent, surrounded by an otherwise flat band with quasienergy $\phi_m$. We choose width $w = 1$, small enough such that the quasienergy is constant with $\mathbf{k}$ at the beginning and end times $t_1$, $t_2$, so as to match the preceding and proceeding evolution.

3. For times $t_1 + t_2 \le t < T - t_1 + t_2$ the eigenvectors remain described by Eq. (4) of the main text at $m = 2$, and the quasienergy increases linearly from $\phi_m$ to $\pi - \phi_m$. This has instantaneous Hopf invariant 1 at all times.

4. For times $(T - t_2) \le t < (T - t_1)$ the process of the second step is replicated, but with the reversed quasienergy $\phi \to \pi - \phi$ and time $t \to T - t$. This gives a gap closing at $\pi$ with charge $-1$.

5. For times $T - t_1 \le t$ the evolution of Fig. 2 is constant. For the micromotion operator, we take the eigenvectors to remain constant (described by Eq. (4) of the main text at $m = 2$), while the quasienergy linearly decreases from $\pi - \phi_m$ to 0. This is homotopy equivalent to the usual micromotion operator $U_m(\mathbf{k}, t) \equiv$

$U(\mathbf{k}, t) \left( U(\mathbf{k}, T) \right)^{-t/T}$ via the deformation $U_m(\mathbf{k}, t) \equiv U(\mathbf{k}, t) \left( U(\mathbf{k}, T) \right)^{-p(t; \tau)}$, where the power $p(t; \tau)$ varies as

$$p(t; \tau) = \begin{cases} 0 & t < \tau \\ (t - T + \tau)/\tau & t \geq \tau \end{cases}, \tag{20}$$

which interpolates between the usual micromotion operator at $\tau = 0$ and our micromotion operator at $\tau = t_1$.

Fig. 2 depicts an evolution with parameters $t_1 = .2T, t_2 = .4T, \phi_m = 3\pi/8$. The phase transition and the smooth deformation were implemented with $t_1 = .3T, t_2 = .35T, \phi_m = .025\pi$ and $t_1 = .3T, t_2 = .35T, \phi_m = .025\pi$, respectively. This evolution has a 0-defect with charge 1 at time $t_0 = (t_1 + t_2)/2$ and a $\pi$-defect with charge $-1$ at time $t_\pi = T - (t_1 + t_2)/2$. A phase transition is naturally found by truncating the evolution at times $T(\lambda) < T$. If the truncation time $T(\lambda)$ occurs after the time of the $\pi$-defect, the system is in the initial phase $(h_S, h_F) = (0, 1)$. However, if the truncation time is *before* the $\pi$-defect, the evolution consists only of the charge 1 0-defect, and thus is in the $(h_S, h_F) = (1, 0)$ phase. The phase transition between the two is at $T(\lambda) = t_\pi$. For Fig. 3, we take $T(\lambda) = T - 2\lambda(T - t_\pi)$, which has a phase transition at $\lambda = 1/2$. The smooth deformation is implemented as described in the main text, with $\theta(t)$ chosen to be the piecewise linear ramp

$$\theta(t) = \begin{cases} (t/t_1)\pi & t < t_1 \\ \pi & t \geq t_1 \end{cases}. \tag{21}$$

The integral for the Floquet Hopf invariant is evaluated by summing the integrand on $N^4$ points in the Floquet Brillouin zone. The points are chosen by first forming an evenly-spaced $N^4$ grid in the Floquet Brillouin zone, and then randomizing the position of each point within the width$-1/N$ hypercube surrounding each grid point. The randomization improves convergence of the integral, and reproduces the non-randomized result in the limit of large $N$. The eigenvectors [of Eq. (4) in the main text] and the phases $\phi(\mathbf{k}, t)$ obey a number of reflection symmetries such that the Floquet invariant integrand is invariant under $k_i \to -k_i$ for any $k_i$. This is used to reduce the domain of the integral to the $0 \leq k_i < \pi$ octant of the Brillouin zone. The $N^4$ grid is shifted by half a discretization in each dimension to avoid over-counting the contribution of the integration domain's boundary. At $N = 90$, the numerical invariants are within 0.6% of their theoretical values on both sides of the phase transition, and within 1.5% of the theoretical value throughout the smooth deformation. As expected, the error is seen to decrease with increasing $N$.

The defect charges, and the static Hopf invariant, are calculated by evaluating the instantaneous Hopf invariants $h(T_{\text{mid}})$ and $h(T)$ at two time-slices of the evolution: an intermediate time $T_{\text{mid}}$ between the two gapless points, and the final time $T$. If the 0-defect occurs before the $\pi$-defect, the defect charges follow as $h_0 = h(T_{\text{mid}})$ and $h_\pi = h(T) - h(T_{\text{mid}})$, and vice versa if the $\pi$-defect occurs before the 0-defect. The static invariant is by definition $h_S = h(T)$. The instantaneous Hopf invariants are numerically calculated similar to the Floquet Hopf invariant, now on a $N^3$ grid in the (non-Floquet) Brillouin zone.

## ADDITIONAL DISCUSSION OF THE EDGE MODES

We briefly review the properties of edge modes in the static HI, and use this to establish the edge modes of the Floquet Hopf insulator in more detail. Unlike TIs classified by K-theory, the edge modes of the HI are protected only for sufficiently smooth edges [4, 5]. This is a consequence of the two-band requirement: the notion of bands relies on translational invariance, which is broken at a sharply terminated edge. At edges that are smooth relative to the lattice length, we may define the adiabatic Hamiltonian $H(\mathbf{k}; \mathbf{r})$. At a boundary where this Hamiltonian's topological invariant changes, the HI's nontrivial homotopy classification requires the Hamiltonian's gap to close, which we identify as the gapless edge mode.

A similar argument holds for the Floquet Hopf insulator, if we replace the adiabatic Hamiltonian with an adiabatic unitary evolution $U(\mathbf{k}, t; \mathbf{r})$. The defect charges $h_0/h_\pi$ of the adiabatic evolution can change by two mechanisms: a $0/\pi$-gap closing, respectively; or the smooth deformation that takes $(h_0, h_\pi) \to (h_0 + 2n, h_\pi - 2n)$. The latter cannot change the charges' parities, which leads to a simple rule: if the parity of $h_0/h_\pi$ changes across a smooth boundary, there will be a gapless edge mode at $0/\pi$-quasienergy. This behavior is identical to that of a stable $\mathbb{Z}_2 \times \mathbb{Z}_2$ Floquet topological insulator.

The more interesting case is that described in the main text, at a boundary where the defect charges' parities do not change, but the charges themselves do (*i.e* the Floquet invariant does not change, but the static invariant changes by an even amount). While the change in static invariant necessitates a gap closing, neither gap individually requires one, and so the quasienergy of the gapless edge mode will be determined by details of the edge region. This is illuminated by a specific example: consider a boundary between the trivial phase, $(h_S, h_F) = (0, 0)$ [e.g. $(h_0, h_\pi) = (0, 0)$], and the phase $(h_S, h_F) = (2, 0)$ [e.g. $(h_0, h_\pi) = (0, 2)$]. A gap closing at $\pi$-quasienergy suffices to change $h_\pi$ by $+2$ and connect the two phases. However, the latter phase is equivalently described by defect charges $(h_0, h_\pi) = (-2, 0)$, which is connected to the $(0, 0)$ phase by a closing at 0-quasienergy. The 'natural' quasienergy of the gapless edge modes depends on the specific defect charges of the two phases, which is a non-universal property of the adiabatic edge region.

---



\end{document}